\documentclass[letterpaper]{article} 
\usepackage{aaai25}  
\usepackage{times}  
\usepackage{helvet}  
\usepackage{courier}  
\usepackage[hyphens]{url}  
\usepackage{graphicx} 
\usepackage{natbib}  
\usepackage{caption} 
\usepackage{algorithm}
\usepackage{algorithmic}
\usepackage{amsmath}
\usepackage{amssymb}
\usepackage{booktabs}
\usepackage{makecell}

\usepackage{newfloat}
\usepackage{listings}
\DeclareCaptionStyle{ruled}{labelfont=normalfont,labelsep=colon,strut=off} 
\floatstyle{ruled}
\newfloat{listing}{tb}{lst}{}
\floatname{listing}{Listing}
\title{Structured  IB: Improving Information Bottleneck with Structured Feature Learning}
\author{%
Hanzhe Yang,
Youlong Wu\thanks{Corresponding authors.},
Dingzhu We, 
Yong Zhou, 
Yuanming Shi\textsuperscript{\rm  *}, 
}
\affiliations {
  School of Information Science and Technology, ShanghaiTech University, Shanghai, China\\
  \{yanghzh2022, wuyl1, wendzh, zhouyong, shiym\}@shanghaitech.edu.cn
}

\usepackage{bibentry}
\newtheorem{theorem}{Theorem}

\begin{document}

\maketitle

\begin{abstract}
The Information Bottleneck (IB) principle has emerged as a promising approach for enhancing the generalization, robustness, and interpretability of deep neural networks, demonstrating efficacy across image segmentation, document clustering, and semantic communication. Among IB implementations, the IB Lagrangian method, employing Lagrangian multipliers, is widely adopted. While numerous methods for the optimizations of IB Lagrangian based on variational bounds and neural estimators are feasible, their performance is highly dependent on the quality of their design, which is inherently prone to errors.
 To address this limitation, we introduce Structured IB, a framework for investigating potential structured features. By incorporating auxiliary encoders to extract missing informative features, we generate more informative representations. Our experiments demonstrate superior prediction accuracy and task-relevant information preservation compared to the original IB Lagrangian method, even with reduced network size.
\end{abstract}

%
\begin{links}
     \link{Code}{https://github.com/HanzheYang19/Structure-IB}
\end{links}

\section{Introduction}
The Information Bottleneck (IB) principle, a representation learning approach rooted in information theory, was introduced by \cite{tishby2000information}. The core idea of IB is to extract a feature, denoted as \(Z\), that captures relevant information about the target \(Y\) from the input \(X\). Specifically, IB aims to maximize the mutual information (MI) between \(Z\) and \(Y\) while constraining the MI between \(X\) and \(Z\), effectively preserving only information essential for predicting \(Y\). Mathematically, this is formulated as:
\begin{equation}
\max_{p(Z|X)} I(Z, Y)\;\;\;\text{s.t.}\;\;I(X, Z)<r,
\end{equation}
where \(r\) controls the level of compression.

Solving this constrained optimization problem directly is challenging. To address this, IB Lagrangian methods (see, \cite{tishby2000information, yu2024cauchy, rodriguez2020convex, shamir2010learning}) introduce a positive Lagrange multiplier \(\beta\), transforming the problem into:
\begin{equation}
\min_{p(Z|X)} -I(Z, Y)+\beta I(X, Z),
\end{equation}
where \(\beta\) is typically within the range [0, 1] \cite{rodriguez2020convex}.

IB and its variations \cite{alemi2016deep, kolchinsky2019nonlinear, rodriguez2020convex} have been applied to various domains, including image segmentation \cite{bardera2009image}, document clustering \cite{slonim2000document}, and semantic communication \cite{yang2023multi, xie2023robust}. Recent research has linked IB to Deep Neural Networks (DNNs) in supervised learning, where \(X\) represents input features, \(Y\) is the target (e.g., class labels), and \(Z\) corresponds to intermediate latent representations. Studies \cite{shwartz2017opening, lorenzen2021information} have shown that IB can explain certain DNN training dynamics. Additionally, IB has been shown to improve DNN generalization \cite{wang2021pac} and adversarial robustness \cite{alemi2016deep}.

Despite their practical success, IB methods face computational challenges during optimization. Variational bounds and neural estimators \cite{belghazi2018mine, pichler2022knife} have been proposed to mitigate these issues \cite{alemi2016deep, kolchinsky2019nonlinear, rodriguez2020convex}. However, the approximations inherent in these methods can deviate from the original objective due to the nature of upper or lower bounds and estimators.

This paper introduces the Structured Information Bottleneck (SIB) framework for IB Lagrangian methods. Unlike traditional IB, which uses a single feature, SIB explores structured features. We divide the feature extractor into a main encoder and multiple auxiliary encoders. The main encoder is trained using the IB Lagrangian, while auxiliary encoders aim to capture information missed by the main encoder. These features are combined to form the final feature. Our experiments demonstrate that SIB achieves higher \(I(Z, Y)\) for the same compression level \(I(X, Z)\), improves prediction accuracy, and is more parameter-efficient. We also analyze how auxiliary features enhance the main feature.

 The contributions of our work can be summarized as follows:
\begin{itemize}
    \item \textbf{Proposed a novel SIB framework for IB Lagrangian methods}, departing from the traditional single feature approach by incorporating multiple auxiliary encoders.
    \item \textbf{Developed a novel training methodology for SIB}, involving a two-stage process: initial training of the main encoder using the IB Lagrangian, followed by training auxiliary encoders to capture complementary information while maintaining distinctiveness from the main feature.
    \item \textbf{Demonstrated superior performance of SIB} compared to existing methods in terms of achieving higher mutual information between the compressed representation and the target variable for the same level of compression, as well as improved prediction accuracy and parameter efficiency.
    \item \textbf{Provided a comprehensive analysis of how auxiliary features enhance the main feature}, shedding light on the underlying mechanisms of the proposed framework.
\end{itemize}

\section{Relative Work}\label{sec:relative}
The IB Lagrangian has been extensively studied in representation learning. Several methods have been proposed to optimize it using DNNs. \cite{alemi2016deep} introduced Variational Information Bottleneck (VIB), deriving variational approximations for the IB Lagrangian:
\begin{equation}
I(X, Z)\le\mathbb{E}_{q(x,z)}\log q(z|x)-\mathbb{E}_{q(t)}\log v(t),
\end{equation}
\begin{equation}
I(Z, Y)\ge\mathbb{E}_{p(y,z)}\log q(y|z)+H(Y),
\end{equation}
where \(q(\cdot|\cdot)\) and \(p(\cdot|\cdot)\) represent the probabilistic mapping and variational probabilistic mapping respectively, \(q(\cdot)\) is the probabilistic distribution, \(H(\cdot)\) is the entropy, and \(v(\cdot)\) is some prior distribution.

Departing from VIB, Nonlinear Information Bottleneck (NIB) by \citet{kolchinsky2019nonlinear} employs kernel density estimation \cite{kolchinsky2017estimating} to bound \(I(X, Z)\):
\begin{equation}I(X, Z)\le-\frac{1}{N}\sum_{i=1}^{N}\log \frac{1}{N}\sum_{j=1}^{N}e^{-D_{KL}[q(z|x_i)\|q(z|x_j)]},\end{equation}
where \(D_{KL}(\cdot\|\cdot)\) is the KL divergence and \(N\) is the total number of samples in the dataset.

However, 
\citet{rodriguez2019information} demonstrated that optimizing the IB Lagrangian for different values of \(\beta\) cannot explore the IB curve when \(Y\) is a deterministic function of \(X\). To address this, they proposed the Square IB (sqIB) by simply using the square of compression term \((I(X, Z))^2\) instead of \(I(X, Z)\) in the Lagrangian function. Furthermore, \citet{rodriguez2020convex} showed that applying any monotonically increasing and strictly convex functions on \(I(X, Z)\) is able to explore the IB curve. Additionally, the authors of \citet{yu2024cauchy} extended the IB for regression tasks with Cauchy-Schwarz divergence.

There are also other approaches that focus on neural estimators for directly approximating MI, entropy, and differential entropy (see, \cite{belghazi2018mine, pichler2022knife}), enabling direct IB Lagrangian optimization without variational bounds.

While these methods offer theoretical advancements, their implementations can be error-prone due to specific implementation details.

Our proposed SIB framework introduces auxiliary encoders to capture complementary information and a novel training process. Unlike prior work, we leverage the IB Lagrangian for both main and auxiliary encoder training and include an additional term to prevent feature overlap.

\section{Methodology}\label{sec:methodolagy}
In this section, we first present the architecture of the proposed network. Subsequently, the training process will be illustrated in detail.

\subsection{Network Architecture}\label{sec:network}
\begin{figure}
  \centering
  \centerline{\includegraphics[width=1\columnwidth]{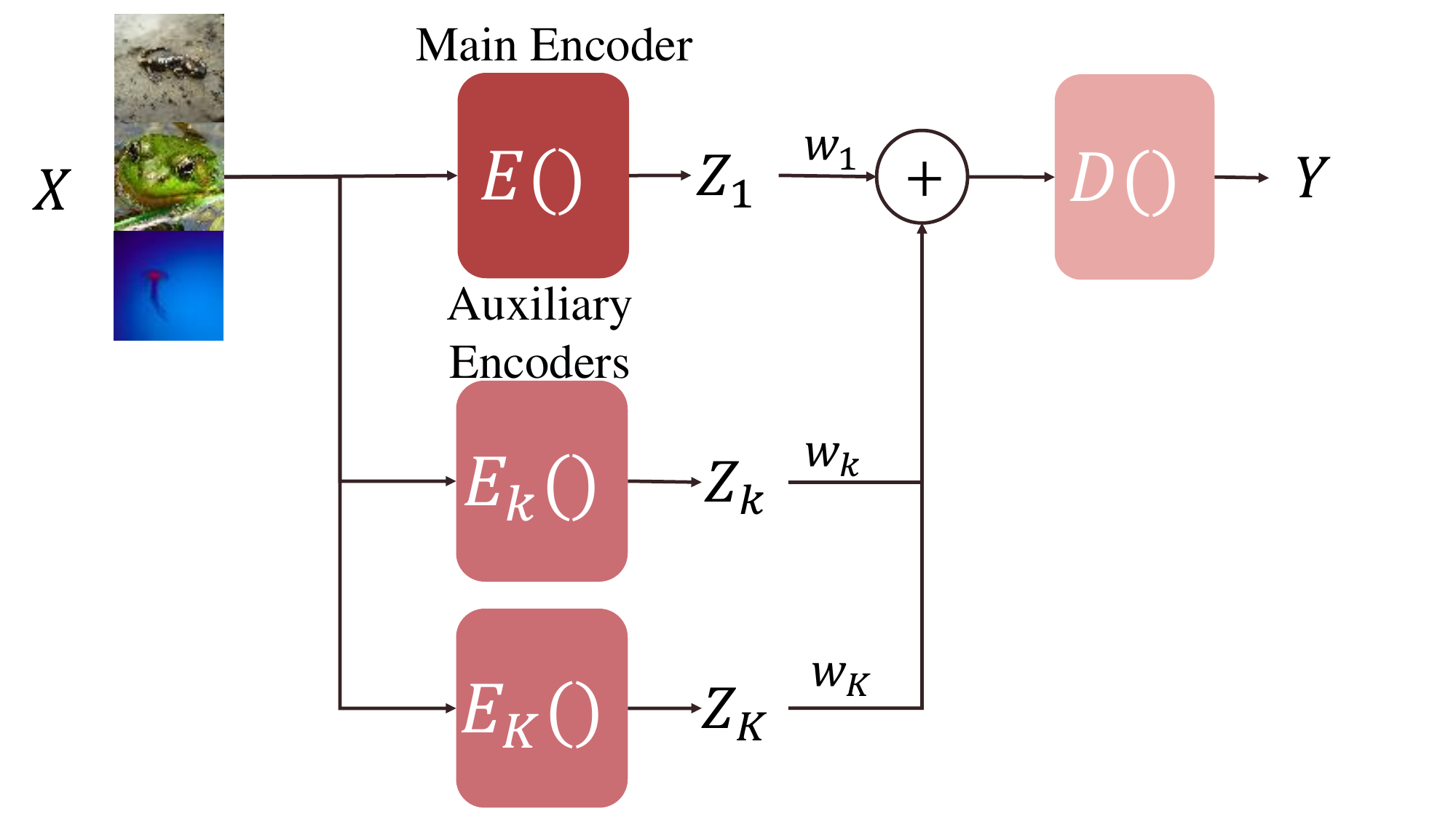}}
  \caption{The illustration of the network architecture.}
  \label{fig:network}
\end{figure}

Our network architecture comprises three key components: the main encoder, auxiliary encoders, and the decoder, as illustrated in Figure \ref{fig:network}.
The main encoder extracts the primary feature from the input, while auxiliary encoders capture complementary features to enrich the information content of the primary feature. These features are combined using a weighted aggregation function \(f(\cdot)\) to produce a comprehensive feature representation. In this work, we employ a simple weighted summation, which will be justified in the following subsection. The decoder processes the aggregated feature to generate the final output.

Let \(X\) and \(Y\) represent the input and label, respectively, with joint distribution \(p(X, Y)\). The main encoder \(E\) extracts the primary feature \(Z = E(X)\), while auxiliary encoders \(E_i\), where \(i = 1, 2, ..., K\) and \(K\) is the number of auxiliary encoders, extract supplementary features \(Z_i = E_i(X)\). These features are weighted by \(w_i, i = 0, 1, ..., K\) and aggregated as follows:
\begin{equation}\hat{Z} = w_0Z+\sum_{i=1}^Kw_iZ_i,\end{equation}
where \(\hat{Z}\) denotes the aggregated feature. Ideally, the main feature should carry the most weight (\(w_0\) is largest). However, we do not enforce this constraint explicitly, as our experiments show that \(w_0\) naturally converges to the dominant weight during training. The decoder \(D\) takes the aggregated feature \(\hat{Z}\) as input and produces the output \(\hat{Y} = D(\hat{Z})\).

\subsection{Training Process}\label{sec:training}
The training process is divided into three stages: training the main encoder and decoder, training auxiliary encoders, and optimizing weights.

\textbf{The main encoder and the decoder:} 
The training of the main encoder and decoder follows traditional IB methods like VIB \cite{alemi2016deep} or NIB \cite{kolchinsky2019nonlinear}. The main encoder \(E\) processes the input \(X\) to generate the primary feature \(Z\), while the decoder \(D\) directly receives \(Z\) to produce a preliminary output \(Y_0\). This forms a Markov chain \(Z \leftrightarrow X \leftrightarrow Y_0\) \cite{witsenhausen1975conditional, gilad2003information}. The goal is to optimize \(E\) to learn the optimal probabilistic mapping \(p(Z|X)\) and \(D\) to approximate the conditional distribution \(p(Y|Z)\) accurately. This is achieved by maximizing the MI \(I(Z, Y)\) while constraining \(I(X, Z)\) through minimizing the IB Lagrangian \cite{gilad2003information, rodriguez2020convex, yu2024cauchy}:
\begin{equation}\min_{E, D} L[p(Z|X), p(Y|Z); \beta] = -I(Z, Y)+\beta I(X, Z),\end{equation}
where \(\beta\) denotes the Lagrange multiplier that balances the trade-off.

\textbf{The auxiliary encoders:} Auxiliary encoders are trained sequentially. For the \(i\)-th encoder \(E_i\) \((i \in [1, K])\), the objective is to extract features that are informative about the target \(Y\) but independent of previously extracted features. 
To capture informative content, we employ the IB Lagrangian:
\begin{equation}\min_{E_i}  L[p(Z_i|X), p(Y|Z_i); \beta] = -I(Z_i, Y)+\beta I(X, Z_i).\end{equation}
Notably, to ensure compatibility and seamless integration of features extracted by different encoders, we maintain a consistent feature space throughout the network. This is achieved by fixing the decoder's parameters after its initial training with the main encoder. Consequently, the decoder remains unchanged during the training of auxiliary encoders. 

To ensure feature independence, we minimize the MI between the current feature \(Z_i\) and the concatenation of previous features \(Z+\sum_{j=1}^{i-1}Z_j\): 
\begin{equation}\min_{E_i}  I(Z_i, Z+\sum_{j=1}^{i-1}Z_j).\end{equation}

Directly minimizing the mutual information \(I(Z_i, Z+\sum_{j=1}^{i-1}Z_j)\), equivalent to the KL divergence \(D_{KL}[p(Z_i, Z+\sum_{j=1}^{i-1}Z_j)|p(Z_i)p(Z+\sum_{j=1}^{i-1}Z_j)]\), is computationally challenging due to the complex distributions involved, each comprising a mixture of numerous components \cite{pan2021disentangled}. To address this, we adopt a sampling-based approach.

First, we generate samples from the joint distribution \(p(Z_i, Z+\sum_{j=1}^{i-1}Z_j)\) by randomly selecting inputs \(X\) from the dataset and extracting the corresponding features. Subsequently, we obtain samples from the product distribution \(p(Z_i)p(Z+\sum_{j=1}^{i-1}Z_j)\) by shuffling the samples from the joint distribution along the batch dimension \cite{belghazi2018mutual}. 

To estimate the KL divergence, we employ the density-ratio trick \cite{nguyen2007estimating, kim2018disentangling}. A discriminator \(d\) is introduced to distinguish between samples from the joint and product distributions. The discriminator is trained adversarially using the following objective:
\begin{equation}
\begin{aligned}
\min_{E_i}\max_{d} \mathbb{E}_{p(Z_i)p(Z+\sum_{j=1}^{i-1}Z_j))}\log d(Z_i, Z+\sum_{j=1}^{i-1}Z_j) +\\
\mathbb{E}_{p(Z_i, Z+\sum_{j=1}^{i-1}Z_j))}\log (1-d(Z_i, Z+\sum_{j=1}^{i-1}Z_j)).
\end{aligned}
\end{equation}
It should be noted that the \(I(Z_i, Z+\sum_{j=1}^{i-1}Z_j)\) is minimized when the Nash equilibrium is achieved (\citet{goodfellow2014generative}).

\textbf{The weights:} The weights \(w_i\) are optimized to fine-tune the overall network using the IB Lagrangian. The objective is to minimize:
\begin{equation}\min_{\{w_i\}_{i\in[0, K]}} L[p(\hat{Z}|X), p(Y|\hat{Z}); \beta] = -I(\hat{Z}, Y)+\beta I(X, \hat{Z}).\end{equation}
It should be noted that the reason why we use weights to be the fine-tune parameter instead of retraining the decoder is that, retraining the decoder could potentially alter the learned relationships between features and the target variable \(Y\), affecting the calculated MI \(I(Z_i, Y)\). By using weights, we preserve the integrity of the trained features while optimizing their combined representation.

Detailed implementation specifics are provided in the supplementary materials.

\subsection{Justification of \(f(\cdot)\)}\label{section:justification}
Any function that increases \(I(\hat{Z}, Y)\), where \(\hat{Z}=f(Z, [Z_i])\), but does not necessarily increase \(I(X, \hat{Z})\), will be appropriate. Here we only justify the choice of weighted sum.

\begin{theorem}
Assume that \(Z, Z' \in \mathbb{R}^D\) are independent, where \(Z\sim\mathcal{N}(\mu, \Sigma)\),  \(Z'\sim\mathcal{N}(\mu', \Sigma')\), \(D\) is the dimension, \(\mu, \mu'\in\mathbb{R}^D\) are the means, and \(\Sigma, \Sigma'\in\mathbb{R}^{D\times D}\) are the diagonal positive definite covariance matrices. Moreover, let \(d(Z)=WZ: \mathbb{R}^D\rightarrow \mathbb{R}^D\) be the linear decoder function with full-rank parameter \(W\in\mathbb R^{D\times D}\), \(h()\) is the one-hot coding function , and \(Y'=d(Z+Z')\). Then, given data \(X\) and its target \(Y\), we have
\begin{equation}I(Z'+Z, Y)\ge I(Z, Y),\end{equation}
when the following conditions are satisfied:
\begin{equation} I(h(Y'), Y)= H(Y),\end{equation}
\begin{equation}
    \det (\Sigma')\ge \frac{1}{(2\pi e)^{D}}.
\end{equation}\label{Theorem}
\end{theorem}
Theorem \ref{Theorem} posits that auxiliary features can augment the information content of the original feature regarding \(Y\) by considering \(wZ+\sum_{j=1}^{i-1}w_jZ_j\) as \(Z\) and \(w_iZ_i\) as \(Z'\) within the SIB framework.

To validate the assumptions underlying Theorem \ref{Theorem}, we employ variational encoders as suggested by \citet{alemi2016deep} to satisfy the Gaussianity assumption. The independence condition is enforced through the minimization of \(I(Z_i, Z+\sum_{j=1}^{i-1}Z_j)\). A one-layer MLP decoder with the dimension of \(Z\) matching the number of classes fulfills the decoder function assumption. Achieving global optimality during training ensures that \(I(h(Y'), Y)= H(Y)\) \cite[Proof of Theorem 2]{pan2021disentangled}. The condition \(\det (\Sigma’)\ge \frac{1}{(2\pi e)^{D}}\) typically holds true due to the large number of classes in practical scenarios. Furthermore, when using NIB, the covariance matrix \(\Sigma\) can be treated as a hyperparameter rather than a trainable parameter.

Regarding \(I(X, \hat{Z})\), we will empirically demonstrate that incorporating auxiliary features remains at a similar level in terms of \(I(X, \hat{Z})\) compared to \(I(X, Z)\). More effort on the theoretical analysis of \(I(X, \hat{Z})\) will be put on.

\begin{figure}
  \centering
  \centerline{\includegraphics[width=1\columnwidth]{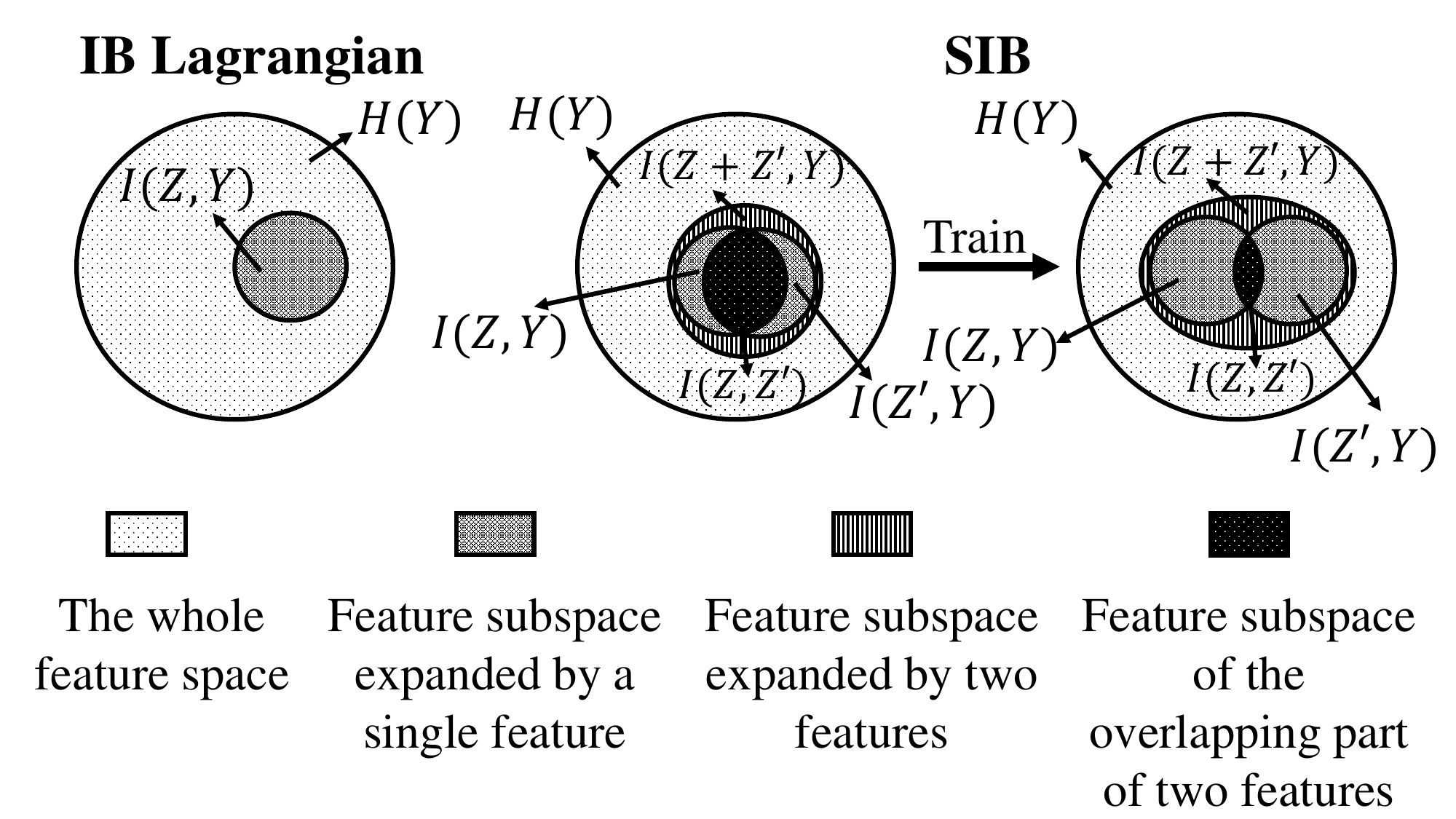}}
  \caption{Illustration of the feature space. \textbf{Left:} The entire feature space (large circle) containing the feature subspace (ellipse) generated by a single feature vector using the IB Lagrangian method.  \textbf{Middle:}  The feature subspace (small circle) spanned by two vectors from an untrained SIB within the overall feature space (large circle). Due to substantial overlap, this subspace is limited in its coverage.  \textbf{Right:} The feature subspace (small circle) is spanned by two vectors from a well-trained SIB within the overall feature space (large circle). Here, the feature subspace is expanded by two minimally overlapping features, maximizing its information content, as demonstrated in Theorem \ref{Theorem}.
}
  \label{fig:illustration}
\end{figure}

We also provide an intuitive explanation of the weighted summation mechanism. Figure \ref{fig:illustration} visually represents the feature spaces of the original IB Lagrangian method and our SIB approach.
In the standard IB Lagrangian method, the encoder generates a single feature, \(Z\), confined to a specific subspace (leftmost circle). While theoretically capable of capturing sufficient information, practical implementations and approximation bounds might lead to information loss.
Our SIB framework introduces an auxiliary feature, \(Z'\), expanding the feature space (middle circle). Initially, the subspaces spanned by \(Z\) and \(Z'\) may overlap significantly. Through the proposed training stages, these subspaces become more distinct as we encourage feature independence \(i.e., (I(Z, Z')<\epsilon\) for a small constant \(\epsilon)\). The subsequent weight optimization further refines the combined subspace spanned by \(Z\) and \(Z'\) to maximize information about \(Y\) as shown in the rightmost circle.

While the weighted aggregation employed in this work effectively combines features, it is acknowledged that a more sophisticated operator capable of fully capturing the intricacies of the feature space could potentially yield even more refined and accurate representations. Future research will explore the development of such operators.

\subsection{Discussion}\label{sec:discussion}
While primarily developed to augment IB Lagrangian methods, our proposed framework is theoretically applicable to any method satisfying the conditions outlined in Theorem \ref{Theorem}. Adapting this framework to other methods would primarily involve modifying the encoders and incorporating the independence-encouraged component into the respective loss function. However, due to the prevalence of IB Lagrangian approaches and computational constraints, our current research focuses on IB Lagrangian implementation.

\section{Experiments}\label{section:experiments}
In this section, we will demonstrate our experimental setups and several results. Specifically, we first evaluate and analyze the performance of the proposed SIB in comparison with its corresponding IB algorithm. Next, we examine the behavior of weights in the SIB, followed by an analysis of both SIB and corresponding IB methods on the IB plane. Due to space limitations, experiments on encoder dropout, which show the significance of SIB weights, are provided in the Appendix.

\subsection{Experiment Setups}\label{sec:setup}

\textbf{Dataset:} Several benchmark datasets are used: MNIST (\citet{lecun1998gradient}) and CIFAR10 (\citet{krizhevsky2009learning}). All the results are based on the test set. 

\textbf{Comparing algorithms:} 
Existing variants of the IB Lagrangian, such as VIB, square VIB (sqVIB), and NIB, are compared to the proposed structured counterparts, SVIB, sqSVIB, and SNIB. Additionally, the estimators MINE and KNIFE, along with their structured versions, are evaluated, whose detailed comparisons is provided in the Appendix.

\textbf{Network structure:}
PyTorch \cite{paszke2019pytorch} is employed to implement the algorithms. Network architectures vary based on the specific algorithm and dataset. In particular, a multilayer perceptron (MLP) is used for MNIST, while a convolutional neural network is adopted for CIFAR-10. Detailed network architectures and hyperparameter settings are provided in the Appendix.

\textbf{Devices:} The experiments were conducted on a system equipped with a 12th Generation Intel(R) Core(TM) i7-12700 CPU, operating at a frequency of 2100MHz. The CPU has 12 cores and 20 logical processors. The system also includes an NVIDIA GeForce RTX 4070 Ti GPU with a WDDM driver and a total available GPU memory of 12282MB.

\subsection{Performance Analysis and Evaluation}\label{sec:performance}
\begin{figure}[t]
  \centering
  \centerline{\includegraphics[width=1.04\columnwidth]{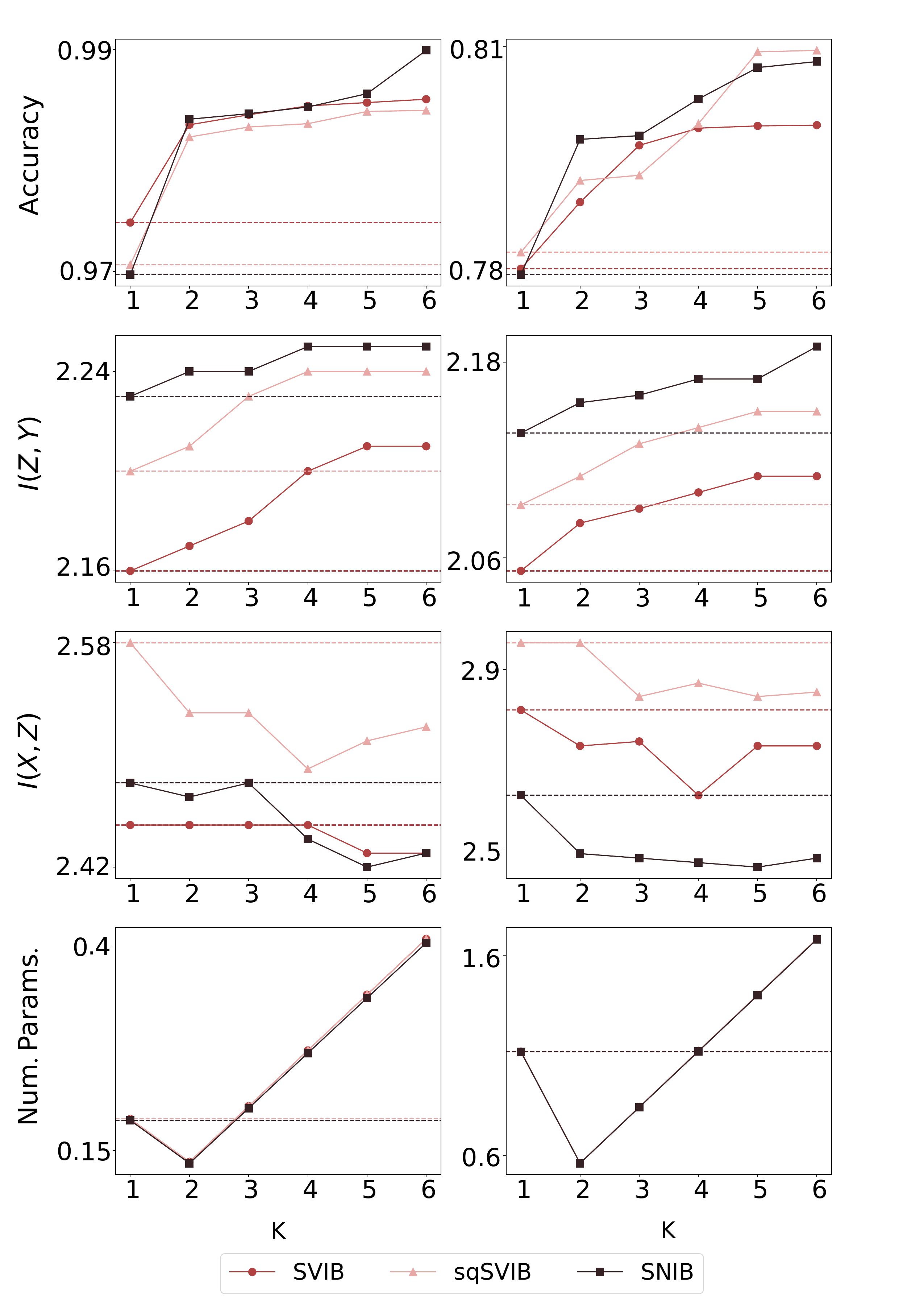}}
  \caption{The accuracy, MI \(I(Z, Y), I(X, Z)\) and the numbers of model parameters (Num. Params., in Millions) v.s. the number of encoders. The IB Lagrangian versions are marked by dashed lines. The left four figures are based on MNIST while the figures in the right are the results of CIFAR10.}
  \label{fig:four}
\end{figure}

The accuracy, \(I(Z,Y), I(X, Z)\), and the number of model parameters with respect to the number of encoders are depicted in Figure \ref{fig:four}. The Lagrange multiplier \(\beta\) is set to \(1\) for SVIB and sqVIB, and \(0.01\) for SNIB. Notably, when \(K=1\), the algorithms reduce to the standard IB Lagrangian. Furthermore, to estimate \(I(X, Z)\), we employ Monte Carlo sampling (\citet{goldfeld2018estimating}). Besides, given that \(I(Z, Y)=H(Y)-H(Y|Z)\), we estimate the conditional entropy \(H(Y|Z)\) using the cross-entropy loss. It is also important to note that \(H(Y)\) is a known constant determined by the dataset. Indeed, as depicted in Figure \ref{fig:four}, the upper four diagrams represent results based on the MNIST dataset, while the lower four diagrams are derived from the CIFAR10 dataset. 

Figure \ref{fig:four} illustrates the accuracy, \(I(Z,Y), I(X, Z)\), and model parameter count in relation to the number of encoders. The Lagrange multiplier, \(\beta\), is set to \(1\) for SVIB and sqVIB, and \(0.01\) for SNIB. Importantly, the algorithms reduce to the standard IB Lagrangian when \(K = 1\). MI \(I(X, Z)\) is estimated using Monte Carlo sampling \cite{goldfeld2018estimating}, while \(I(Z, Y)=H(Y)-H(Y|Z)\) is derived from the conditional entropy \(H(Y|Z)\), calculated via cross-entropy loss and a dataset-dependent constant \(H(Y)\). The upper and lower rows of Figure \ref{fig:four} present results for MNIST and CIFAR-10, respectively.

For MNIST dataset,
our structured algorithms consistently outperform the original IB Lagrangian in terms of accuracy and \(I(Z,Y)\) even when the number of model parameters is reduced. This improvement is attributed to our method's ability to extract supplementary features that enhance the information content of the main feature. However, as \(K\) increases, the performance may meet the bottleneck which will prevent it from surging. Thus, the number of encoders which determine the amount of parameters should be chosen carefully. In most cases, \(I(X, Z)\) experiences a slight decrease as well. However, the mechanism behind this phenomenon should be investigated in the future. The performance on the CIFAR10 dataset is similar to that of MNIST.

For the MNIST dataset, our structured algorithms consistently surpass the original IB Lagrangian in terms of accuracy and \(I(Z,Y)\), even with reduced model parameters. This enhancement is attributed to our method's capacity to extract complementary features that enrich the primary feature's information content. Nevertheless, increasing the number of encoders \(K\) may lead to performance plateaus, necessitating careful selection of this hyperparameter. While \(I(X, Z)\) generally exhibits a slight decline, the underlying cause of this behavior warrants further exploration. Notably, the CIFAR-10 dataset demonstrates performance trends similar to MNIST.

In summary, our proposed method consistently outperforms traditional approaches in terms of accuracy and feature extraction, even when using smaller network architectures. To optimize performance and computational efficiency, careful selection of the hyperparameter \(K\) is crucial, as diminishing returns may occur with excessively large values.
\begin{table*}[t]
  \centering
  \begin{tabular}{llll}
    \toprule
    \multicolumn{3}{r}{MNIST}                   \\
    \cmidrule(r){2-4}
    K     & SVIB     & sqSVIB & NIB \\
    \midrule
    2 & (0.9453, 0.0833)  & (0.9302, 0.1041) &(0.9498, 0.0907)    \\
    3     & (0.8901, 0.0441, 0.1159) & (0.9202, 0.0407, 0.0629)&(0.7927, 0.0729, 0.1438)     \\
    \midrule
    \multicolumn{3}{r}{CIFAR10} \\
    \cmidrule(r){2-4}
    & SVIB     & sqSVIB & NIB \\
    \midrule
    2& (0.8061, 0.2383)&(0.9345, 0.0872)&(0.8323, 0.2574)\\
    3& (0.8037, 0.0889, 0.1659)&(0.6684, 0.2058, 0.2626)&(0.8457, 0.2040, 0.1275)\\

    \bottomrule
  \end{tabular}
\caption{The weights of encoders.}\label{table:w}
\end{table*}

\subsection{Behavior of Weights}\label{sec:weights}
In this experiment, the Lagrange multiplier \(\beta\) was set to 1 for SVIB and sqVIB, and 0.01 for SNIB. Additionally, the achieved accuracies consistently matched those reported in the previous subsection. 

Intriguingly, we observed a consistent pattern where the main encoder was assigned the largest weight, even without explicit constraints. As detailed in Table \ref{table:w}, the weights collectively approximated 1 despite lacking explicit normalization. Furthermore, normalizing the weights to sum exactly to one had negligible impact on performance metrics. Consequently, additional weight manipulations were deemed unnecessary.

Furthermore, as detailed in Table \ref{table:w}, the auxiliary encoder weights are generally larger for the CIFAR-10 dataset compared to MNIST, indicating that these encoders contribute more significantly to the overall performance on more complex tasks.

\subsection{Behavior on IB Plane}\label{sec:IBP}
\begin{figure}[t]
  \centering
  \centerline{\includegraphics[width=1.05\columnwidth]{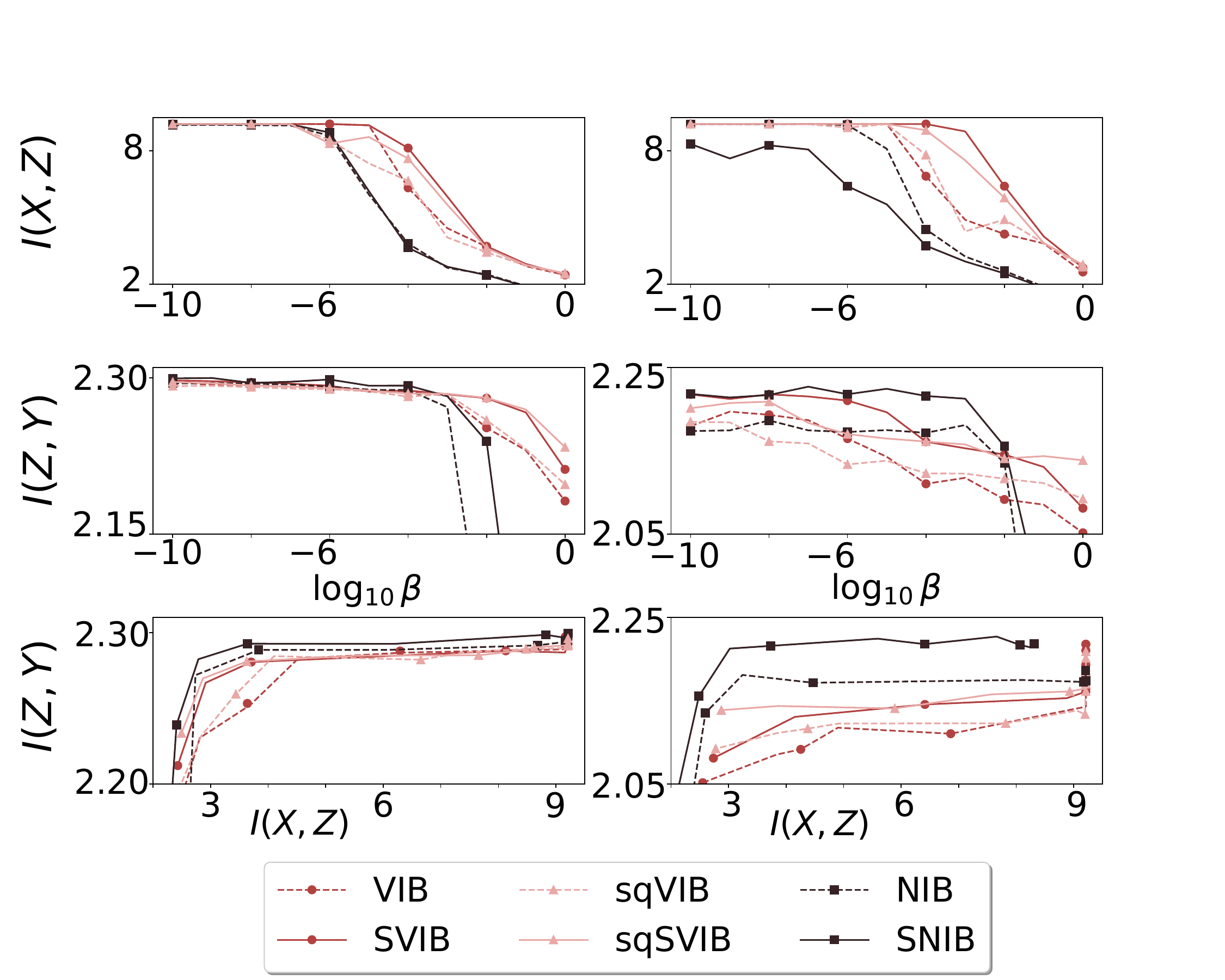}}
  \caption{Behavior of Algorithms on the IB Plane. The original IB Lagrangian methods and their structured counterparts are represented in the same color, differentiated by solid and dashed lines. The left and right figures correspond to the MNIST and CIFAR10 datasets, respectively.}
  \label{fig:IB}
\end{figure}

This section analyzes the behavior of our proposed algorithms and original IB Lagrangian algorithms on the IB plane, using two encoders, i.e., \(K=2\). Figure \ref{fig:IB} presents results for MNIST (left column) and CIFAR-10 (right column) datasets. Our methods consistently achieve higher \(I(Z, Y)\) compared to the original IB Lagrangian. However, the rate of decrease in \(I(X, Z)\) is generally slower. Notably, SNIB consistently exhibits lower \(I(X, Z)\) than NIB. Furthermore, the \(I(X, Z)\) of SVIB and sqSVIB converges to that of VIB and sqVIB as the Lagrange multiplier \(\beta\) increases. Crucially, our methods demonstrate the ability to attain higher \(I(Z, Y)\) for a given level of \(I(X, Z)\) as revealed in the third row.

In summary, our proposed method effectively improves the informativeness of extracted features while maintaining or even enhancing data compression (especially for SNIB). Additionally, our models achieve superior feature utility at comparable compression levels compared to the original IB Lagrangian. Significantly, these advantages are realized with a reduced number of model parameters as shown in Figure \ref{fig:four}.



\section{Conclusion and Discussion}\label{sec:conclusion}
This paper introduces a novel structured architecture designed to improve the performance of the IB Lagrangian method. By incorporating auxiliary encoders to capture additional informative features and combining them with the primary encoder's output, we achieve superior accuracy and a more favorable IB trade-off compared to the original method, even with a significantly smaller network. Our results highlight the effectiveness of our approach in enhancing feature utility while maintaining desired compression levels. Furthermore, we demonstrate the crucial role of auxiliary encoders in driving these improvements.

Several limitations warrant further investigation. The current linear aggregation function may oversimplify feature space interactions, necessitating more sophisticated combination strategies. Additionally, theoretical advancements, such as relaxing the Gaussian assumption and investigating how to assist the compression, could refine both the aggregation function and decoder design. While this study focuses on IB Lagrangian methods, our structured architecture holds potential for broader applications. Overall, we believe our proposed framework offers a promising foundation for future research.

\bibliography{aaai25}

\newpage
We provide additional materials that support our main theorem and experiments.

\appendix

\section{Proof of Theorem \ref{Theorem}}\label{appendix}
With \(I(h(Y'), Y) = H(Y)\), we know that \(I(Y', Y)=H(Y)\), since \(H(Y)\ge I(Y', Y)\ge I(h(Y'), Y)\). Thus, \(Y\) can be regarded as a bijective and deterministic function of \(Y'\). Assume this function is denoted as \(f()\). With the Markov chain \(Z\rightarrow (Z+Z')\leftrightarrow Y'\leftrightarrow Y\xrightarrow{a} Y'\), where the arrow \(a\) is because we can further add a node \(Y'\) into this chain by \(Y'=f^{-1}(Y)\). 
Thus \(I(Z, Y)=I(Z, Y')\), as \(I(Z, Y)\ge I(Z, Y')\) and \(I(Z, Y)\le I(Z, Y')\). Similarly, we have \(I(Z+Z', Y)=I(Z+Z', Y')\)

Now, we can directly compare \(I(Z'+Z, Y')\) and \(I(Z, Y')\). We can obtain:
\begin{align}
    I(Z'+Z, Y')-I(Z, Y') &=  H(Z'+Z)-I(Z, Z+Z'), \label{a}\\ 
    &=H(Z'+Z)-H(Z). \label{b}
\end{align}
Here, \eqref{a} is because \(Y'=D(Z+Z')=W(Z+Z')\) is a bijective and deterministic function. Additionally, \eqref{b} is because for any random variables \(X \text{ and } Y\), \(I(X, Y)=H(X)+H(Y)-H(X, Y)\).

Given \( Z\) and \( Z' \) are \(D\)-dimensional independent Gaussian random vectors, we now compare \( H(Z+ Z', Z) \) and \( H(Z) \).

The differential entropy of a \(D\)-dimensional Gaussian random vector \( Z\) with mean vector \( \mu \) and covariance matrix \( \Sigma \) is given by:
\begin{equation}
H(Z) = \frac{1}{2} \log \left( (2\pi e)^D \det(\Sigma) \right),
\end{equation}
where \(\det()\) is the determinant of a matrix. 

To find \( H(Z+ Z', Z) \), we consider the joint distribution of \( (Z+ Z', Z) \).

Since \( Z\sim \mathcal{N}(\mu, \Sigma) \) and \( Z' \sim \mathcal{N}(\mu', \Sigma') \) are independent, \( Z+ Z' \) is also Gaussian:
\begin{equation}
Z+ Z' \sim \mathcal{N}(\mu + \mu', \Sigma + \Sigma').
\end{equation}

Thus, the joint distribution of \( (Z+ Z', Z) \) is a \( 2D \)-dimensional Gaussian with mean vector:
\begin{equation}
\mu''=
\begin{pmatrix}
\mu + \mu' \\
\mu
\end{pmatrix},
\end{equation}
and the covariance matrix:
\begin{equation}
\Sigma'' = \begin{pmatrix}
\Sigma + \Sigma' & \Sigma \\
\Sigma & \Sigma
\end{pmatrix}.
\end{equation}

For a \( 2D \)-dimensional Gaussian distribution with covariance matrix \( \Sigma'' \), the differential entropy is given by:
\begin{equation}
H(Z+ Z', Z) = \frac{1}{2} \log \left( (2 \pi e)^{2D} \det(\Sigma'') \right).
\end{equation}

Using the block matrix determinant formula, the determinant of \( \Sigma'' \) is:
\begin{equation}
\det(\Sigma'') = \det(\Sigma + \Sigma') \cdot \det(\Sigma - \Sigma(\Sigma + \Sigma')^{-1}\Sigma).
\end{equation}

Given that \( \Sigma - \Sigma(\Sigma + \Sigma')^{-1}\Sigma = \Sigma (\Sigma' (\Sigma + \Sigma')^{-1}) \), we can obtain:
\begin{equation}
\det(\Sigma'') = \det((\Sigma+\Sigma')\Sigma\Sigma'(\Sigma+\Sigma')^{-1}) = \det{\Sigma\Sigma'},
\end{equation}
as both \(\Sigma\) and \(\Sigma'\) are diagonal.

Now, we can obtain:
\begin{equation}
H(Z+ Z', Z) = D \log(2 \pi e) + \frac{1}{2} \log (\det(\Sigma\Sigma')),
\end{equation}
and
\begin{equation}
H(Z) =  \frac{D}{2} \log(2 \pi e) + \frac{1}{2} \log (\det(\Sigma)).
\end{equation}

Thus, we have
\begin{equation}
H(Z+ Z', Z)-H(Z) = \frac{D}{2} \log(2 \pi e) + \frac{1}{2} \log (\det(\Sigma')).
\end{equation}

Clearly, 
\begin{equation}
H(Z+ Z', Z) \ge H(Z), 
\end{equation}
when
\begin{equation}
    \det (\Sigma')\ge \frac{1}{(2\pi e)^{D}}.
\end{equation}

This completes the proof.

\section{Implementation Details}\label{appen}
The networks for IB Lagrangian are shown in Table \ref{Table:NS}. For the proposed structured method, we only change the encoders for fair comparison. For fully connected layers, the hidden dimension of each encoder is divided by 2. For example, the encoder \(10\rightarrow 100\rightarrow 16\) will be modified as \(10\rightarrow 100/2\rightarrow 16\). For convolutional layers, the number of channels of each encoder is divided by \(2\). 
\begin{table*}
  
  \centering
  \begin{tabular}{lll}
    \toprule

         & MNIST     & CIFAR \\
    \midrule
    (sq)VIB Encoder &  \makecell[l]{nn.Linear(784, 1024),\\
            nn.ReLU(),\\
            nn.Linear(1024, 1024),\\
            nn.ReLU(),\\
            nn.Linear(1024, 16*2)}  & \makecell[l]{ResNet18 convolution layers \\with fully connected layer nn.Linear(512, 16 *2)}    \\
    \midrule
    NIB Encoder &  \makecell[l]{nn.Linear(784, 1024),\\
            nn.ReLU(),\\
            nn.Linear(1024, 1024),\\
            nn.ReLU(),\\
            nn.Linear(1024, 16)}  & \makecell[l]{ResNet18 convolution layers \\with fully connected layer nn.Linear(512, 16)}   \\
    \midrule
    Discriminator &  \makecell[l]{nn.Linear(16*2, 1000),\\
            nn.ReLU(),\\
            nn.Linear(1000, 2000),\\
            nn.ReLU(),\\
            nn.Linear(2000, 1),\\
            nn.Sigmoid()} & The same as MNIST   \\
    \midrule
    Decoder & \makecell[l]{nn.Linear(16, 10)} & The same as MNIST\\
    \bottomrule
  \end{tabular}
\caption{Network structures}\label{Table:NS}
\end{table*}

Additionally, Adam optimizer is used. The learning rate is set to be 0.0001. We train the algorithm for 200 epochs with a batch size of 100. The Lagrange multiplier \(\beta\) and the number of structured encoders \(K\) are changed through different experiments which have been introduced in the specific subsections.

Moreover, the implementation details of the training algorithm can be found in Algorithm 1 as shown below.

\section{Encoder Dropout}\label{sec:dropout}
\begin{figure*}[t]
  \centering
  \centerline{\includegraphics[width=1.5\columnwidth]{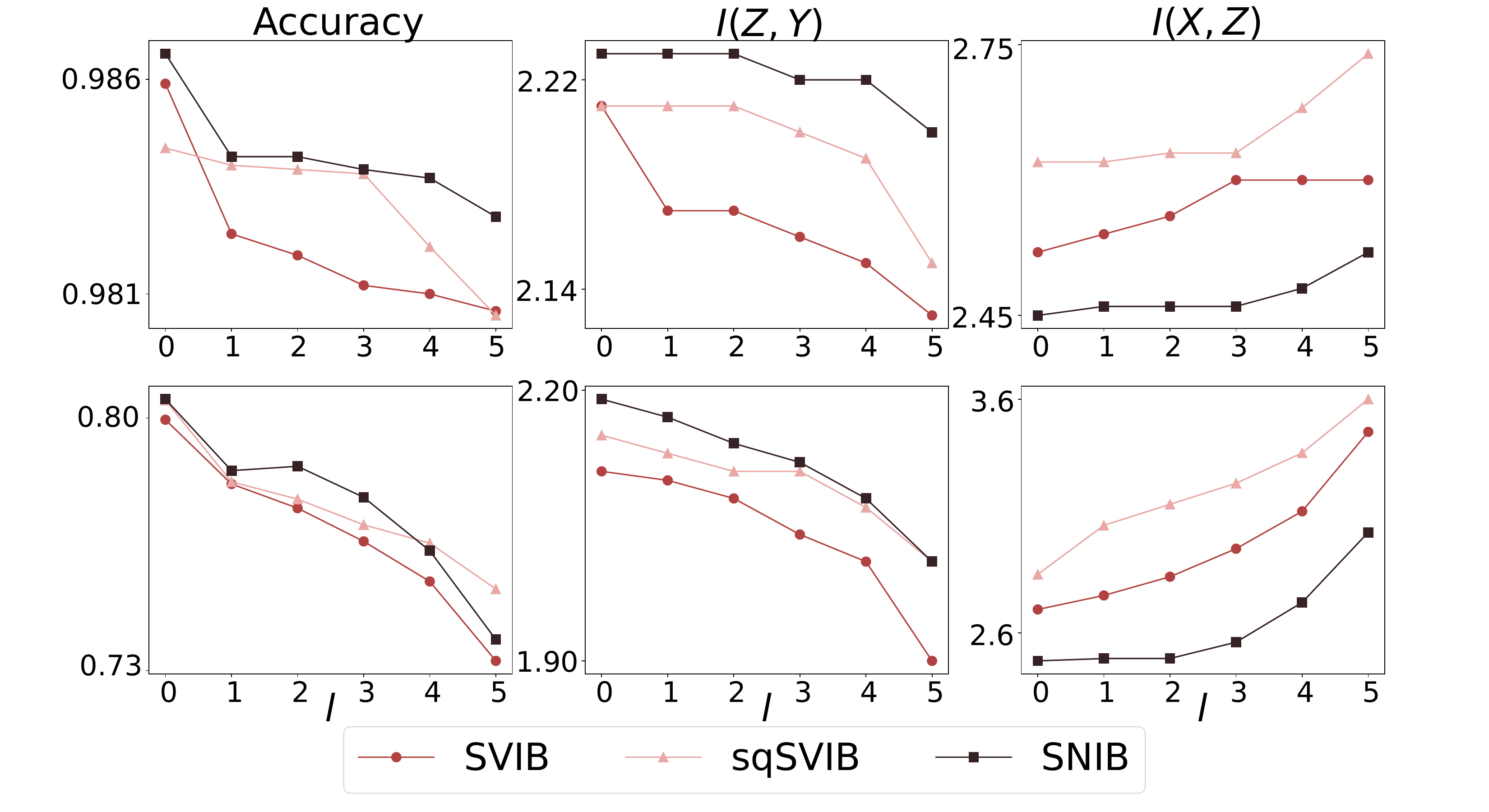}}
  \caption{The performance after encoder dropout. The upper and lower rows correspond to the MNIST and CIFAR10 datasets, respectively.}
  \label{fig:drop}
\end{figure*}

As previously discussed, 
weights play a crucial role in our method. To further investigate the significance of weights and auxiliary encoders, we set \(K=6, \beta=1\) for SVIB and sqVIB, and \(\beta=0.01\) for SNIB. Subsequently, we progressively removed auxiliary encoders with the smallest absolute weight values. Figure \ref{fig:drop} illustrates that both accuracy and \(I(Z, Y)\) consistently decline as more encoders are eliminated. Intriguingly, \(I(X, Z)\) increases with the number of removed encoders. This implies that auxiliary encoders not only carry valuable information but also potentially contribute to compression. Moreover, these findings corroborate the claim in Section \ref{section:justification} that \(I(\hat{Z}, Y)\ge I(Z, Y)\) and \(I(X, \hat{Z})\le I(X, Z)\).

\section{Supplimentary Experiments of Neural Estimators}
This section presents experimental results for MINE and KNIFE. However, the original algorithms, and consequently their structured counterparts, often encountered convergence issues, resulting in unacceptably low accuracy (approaching random guessing) for certain values of \(\beta\). This behavior is probably attributed to the estimation inaccuracy of \(I(X, Z)\), which frequently yielded NaN values. As a result, the corresponding points in the IB plane could not be plotted. Similar issues arose when employing encoder dropout, further emphasizing the importance of auxiliary encoders in stabilizing the training process. Therefore, results for encoder dropout are omitted from this study.
\subsection{MINE}
The Lagrangian parameter \(\beta\) was set to \(1e-7\) to prevent algorithm instability associated with larger values. Given the computational demands for large \(K\), the number of encoders was restricted to the range \([1, 3]\).

For the MNIST dataset, accuracies of \([0.9720, 0.9792, 0.9853]\) and corresponding \(I(\hat{Z}, Y)\) values of \([2.29, 2.29, 2.30]\) were obtained for \(K = 1, 2, 3\) respectively. Notably, \(I(X, \hat{Z})\) remained constant at 9.21 due to the small \(\beta\) value. The number of parameters was consistent with variational bound-based methods as network architectures were unchanged.

A similar trend was observed for the CIFAR-10 dataset, with accuracies of \([0.7659, 0.7738, 0.7821]\) and \(I(\hat{Z}, Y)\) values of \([2.19, 2.20, 2.22]\) for \(K = 1, 2, 3\), respectively. Again, \(I(X, \hat{Z})\) was stable at 9.21.

These findings indicate a modest improvement in the informativeness of the aggregated feature.

\begin{table}
  \centering
  \begin{tabular}{lll}
    \toprule
    \multicolumn{2}{r}{MNIST}                   \\
    \cmidrule(r){2-3}
    K     & SMINE     & SKNIFE  \\
    \midrule
    2 & (1.453, 0.751)  & (1.365, 0.711) \\
    3 & (1.673, 0.548, 0.761) & (1.564, 0.642, 0.870) \\
    \midrule
    \multicolumn{2}{r}{CIFAR10} \\
    \cmidrule(r){2-3}
    & SMINE     & SKNIFE \\
    \midrule
    2& (2.050, 0.424)&(1.834, 0.287)\\
    3& (1.186, -0.583, 1.355)&(2.133, 0.767, 0.431)\\
    \bottomrule
  \end{tabular}
\caption{The weights of encoders.}\label{table:ww}
\end{table}
\subsection{KNIFE}
Using the same \(\beta\) value of \(1e-7\) as for MINE, similar trends were observed for the KNIFE algorithm.

On the MNIST dataset, accuracies of \([0.9776, 0.9838, 0.9862]\) and corresponding \(I(\hat{Z}, Y)\) values of \([2.29, 2.30, 2.30]\) were obtained for \(K = 1, 2, 3\) respectively. Similar to MINE, \(I(X, \hat{Z})\) remained constant at 9.21.

For the CIFAR-10 dataset, accuracies of \([0.7710, 0.7769, 0.7832]\) and \(I(\hat{Z}, Y)\) values of \([2.20, 2.22, 2.22]\) were achieved for \(K = 1, 2, 3\), respectively. Again, \(I(X, \hat{Z})\) was stable at 9.21.

These results align with those obtained for MINE, indicating a consistent improvement in the informativeness of the aggregated feature for both algorithms. 
\subsection{Behaviors of Weights}
While model hyper-parameters remained consistent, a notable discrepancy emerged in weight behavior, as shown in Table \ref{table:ww}. Unlike previous variational bound methods where weights summed to approximately 1, the estimator-based methods exhibited weights accumulating to roughly \(K\). However, the main encoder consistently retained the largest weight in both cases. Moreover, the weights associated with CIFAR-10 tend to be smaller than those for MNIST, suggesting a greater reliance on the main encoder for more challenging tasks. This contrasts with variational bound methods, where the opposite trend is often observed. Interestingly, when the main encoder appears to be underperforming, as observed in SMINE on CIFAR-10 with K=3, auxiliary encoders can contribute significantly to correcting its output. 

\begin{algorithm*}[tb]
\caption{Training algorithm of SIB.}
\label{alg:algorithm}
\textbf{Input}: Training set \(\mathcal{D}=\{x_i, y_i\}_{i=1}^N\), encoders \(E, \{E_i\}_{i=1}^K\), decoder \(D\), discriminator \(d\), weights \(\{w_i\}_{i=0}^K\), \(epoch\) and \(\beta\).\\
\textbf{Output}: Well-trained encoders \(E, \{E_i\}_{i=1}^K\), decoder \(D\) and weights \(\{w_i\}_{i=0}^K\).
\begin{algorithmic}[1] 
\FOR{\(e=1\text{ to } epoch\)}
\FOR{each random batch \(\{x_j, y_j\}_{j\in \mathcal{B}}\)}
\STATE  train \(E\) and \(D\) by \(\min_{E, D} \frac{1}{|\mathcal{B}|}\sum_{j\in \mathcal{B}}-I(E(x_j), y_j)+\beta I(x_j, E(x_j));\)
\ENDFOR
\ENDFOR
\FOR{\(e=1 \text{ to } epoch\)}
\FOR{\(i\in [1, K]\)}
\FOR{each random batch \(\{x_j, y_j\}_{j\in \mathcal{B}}\)}
\STATE Obtaining samples \(\{z_{ij}=E_i(x_j), z_j=E(x_j), z_{lj}=E_l(x_j)\}_{j\in\mathcal{B}, l\in[1, i-1]};\)
\STATE Training \(E_i\) by \(\min_{E_i} \frac{1}{|\mathcal{B}|}\sum_{j\in \mathcal{B}}-I(z_{ij}, y_j)+\beta I(x_j, z_{ij})-\log d(z_{ij}, z+\sum_{l\in[1,i-1]} z_{lj});\)
\STATE Shuffling on batch indices \(\mathcal{B}\), obtaining \(\{z_{ij}, z_{\pi(j)}+\sum_{l\in[1,i-1]} z_{l\pi(j)}\}_{j\in\mathcal{B}}\) where \(\pi()\) is the shuffling projection;\\
\STATE Training \(d\) by \(\min_{d} \frac{1}{|\mathcal{B}|}\sum_{j\in \mathcal{B}}-\log(1-d(z_{ij}, z+\sum_{l\in[1,i-1]} z_{lj}))-\log d(z_{i\pi(j)}, z_{\pi(j)}+\sum_{l\in[1,i-1]} z_{l\pi(j)});\)\\
\ENDFOR
\ENDFOR
\ENDFOR
\FOR{\(e=1 \text{ to } epoch\)}
\FOR{each random batch \(\{x_j, y_j\}_{j\in \mathcal{B}}\)}
\STATE Calculate \(\hat{z}_j=w_0E(x_j)+\sum_{i=1}^K w_iE_i(x_j);\)
\STATE Training weights \(\{w_i\}_{i=0}^K.\) by \(\min_{\{w_i\}_{i=0}^K} \frac{1}{|\mathcal{B}|}\sum_{j\in \mathcal{B}}-I(\hat{z}_j, y_j)+\beta I(x_j, \hat{z}_j);\)
\ENDFOR
\ENDFOR
\STATE \textbf{return} \(E, \{E_i\}_{i=1}^K\), \(D\) and \(\{w_i\}_{i=0}^K\).
\end{algorithmic}
\end{algorithm*}

\end{document}